\documentclass[twocolumn]{aastex701}

\usepackage{amsmath}
\graphicspath{{./}}
\shorttitle{Lower-Mass Giants Spare Inner Planets}
\shortauthors{Jiang}
\submitjournal{ApJL}

\begin{document}

\title{Lower-Mass Giants Can Spare Inner Planets during High-Eccentricity Migration}

\correspondingauthor{Qunfeng Jiang}
\author[orcid=0009-0008-0600-3743,gname=Qunfeng,sname=Jiang]{Qunfeng Jiang}
\affiliation{Independent Researcher, Shanghai, China}
\email[show]{qfjiang01@gmail.com}

\begin{abstract}
The survival of an inner planet next to a close-in giant is commonly used to argue against high-eccentricity migration (HEM), because the eccentric giant is expected to destabilize the inner system. Existing survival calculations have concentrated on Jupiter-mass migrants, although several Jupiter-sized close-in giants with nearby companions have sub-Jovian masses. We use dissipative three-body integrations to measure the collision--survival boundary at seven giant masses from 0.25 to $1.0\,M_{\rm Jup}$ and combine it with uncompressed equilibrium-tide circularization times. At fixed physical periapse, lower giant mass weakens the forcing of the inner orbit and shifts the survival boundary inward. It also shortens the circularization time for a common radius and tidal prescription. These dynamical and tidal limits enclose a viable HEM region in which a sub-Jovian giant can circularize within a system age while retaining an ultra-short-period companion. We screen known companion-hosting close-in giants and perform system-specific integrations of WASP-132; every current candidate lies outside the region. A future system inside it would show directly that HEM can produce a close-in giant while preserving an inner planet, converting companion survival into positive evidence for this formation channel.
\end{abstract}


\section{Introduction}\label{sec:introduction}

Close-in giant planets require substantial orbital evolution after formation. Disk-driven migration can move a giant inward on a nearly circular orbit, whereas high-eccentricity migration (HEM) first excites a wide-orbit planet to a small periapse and then removes orbital energy through repeated tidal passages \citep{Lin1996,RasioFord1996,DawsonJohnson2018}. Scattering, secular interactions, and Lidov--Kozai cycles can supply the eccentricity excitation \citep{WuMurray2003,FabryckyTremaine2007,Nagasawa2008,BeaugeNesvorny2012,Naoz2016}, and the properties of distant giant companions indicate that this pathway contributes to the close-in giant population \citep{PetrovichTremaine2016,ZinkHoward2023}.

An inner companion can test this migration history. During HEM, the giant traverses the inner system at high eccentricity and can scatter a low-mass planet, drive it into the star, or sustain tide-assisted orbital decay \citep{Mustill2015}. The scarcity of nearby planets around hot Jupiters supports this destructive outcome \citep{Steffen2012,Huang2016}. The known exceptions include WASP-47, WASP-132, WASP-84, TOI-1408, TOI-2494, and TOI-5143 \citep{Becker2015,Hord2022,Maciejewski2023,Korth2024,Quinn2026}; transit-timing variations and a uniform census suggest that the population may be larger \citep{WuRiceWang2023,Sha2026}. Interpreting these systems requires a quantitative survival criterion over the relevant giant-planet masses.

Recent dynamical calculations established a survival scale for a Jupiter-mass migrant \citep{Becker2026}. An Earth-mass planet on an ultra-short-period orbit commonly survives when the two orbits remain separated by roughly 14 mutual Hill radii at the giant's periapse, although the observed systems examined in that study lose their inner planets. Several of the known Jupiter-sized close-in giants with nearby companions have sub-Jovian masses. A survival criterion calibrated to a $1\,M_{\rm Jup}$ migrant therefore need not apply uniformly to the observed companion-hosting population. The mass dependence of the physical collision--survival boundary, and therefore of the formation inference drawn from an inner companion, has not been quantified.

Giant mass affects both the perturbation of the inner orbit and the tidal damping time of the eccentric giant. Lower mass may permit companion survival at a smaller periapse, while also changing whether the giant can circularize within the system age. We therefore ask: \emph{does the presence of an inner companion rule out HEM equally strongly at all giant-planet masses?}

We address the question with two linked calculations. Dissipative three-body integrations at fixed inner architecture measure the collision--survival boundary at seven masses across $M_g=0.25$--$1.0\,M_{\rm Jup}$ and $q_g=0.03$--0.06 au. Uncompressed equilibrium-tide calculations then determine where the surviving giant can circularize within 1 or 5 Gyr. The two limits define a viable HEM region, which we compare with the known companion-hosting systems and test directly for WASP-132. Section~\ref{sec:methods} describes the integrations and tidal calculation. Section~\ref{sec:results} measures the two boundaries and their viable region. Section~\ref{sec:applications} evaluates the observed systems, and Section~\ref{sec:discussion} gives the formation implications and limitations.

\section{Numerical Experiments and Tidal Completion}\label{sec:methods}

\subsection{Controlled planetary architecture}\label{sec:architecture}

The reference system is designed to isolate the influence of the migrating giant's mass. A $1\,M_\odot$ star hosts a $1\,M_\oplus$, $1\,R_\oplus$ inner planet on a circular orbit at $a_{\rm in}=0.02$ au. This orbit lies at the outer edge of the ultra-short-period regime around a solar-mass star, where tides and relativistic precession can compete with the perturbation from an eccentric giant. Holding the inner orbit fixed separates the mass dependence studied here from the strong dependence on inner semimajor axis already established for a Jupiter-mass migrant \citep{Becker2026}.

The giant begins at $a_g=2$ au, representing a planet that formed well outside the compact inner system and was subsequently excited to high eccentricity. Its eccentricity is set by the physical periapse,
\begin{equation}
 e_g=1-\frac{q_g}{a_g}.
 \label{eq:eout}
\end{equation}
We use $M_g=0.25$, 0.40, 0.50, 0.60, 0.70, 0.85, and $1.0\,M_{\rm Jup}$, spanning hot-Saturn to hot-Jupiter masses, and set $R_g=1\,R_{\rm Jup}$ in every run so that mass is the controlled variable. Periapses of 0.03--0.06 au cover nearly crossing encounters through well-separated passages; approximate conservation of orbital angular momentum maps this range to final circular orbits near 0.06--0.12 au. The orbits are initially coplanar and $e_{\rm in}=0$, focusing the experiment on the radial encounters and secular forcing that operate even without inclination excitation.

For comparison with earlier stability work, we report the initial spacing at periapse in mutual Hill radii,
\begin{equation}
 \begin{aligned}
 \Delta_H&=\frac{q_g-a_{\rm in}}{R_{H,{\rm peri}}},\\
 R_{H,{\rm peri}}&=\left(\frac{M_{\rm in}+M_g}{3M_\star}\right)^{1/3}
 \frac{a_{\rm in}+q_g}{2},
 \end{aligned}
 \label{eq:hill}
\end{equation}
using the mutual Hill scale for two coplanar planets \citep{Gladman1993}. The main comparison fixes $q_g$, so giants of different masses pass through the same physical location relative to the inner planet.

The numerical program proceeds from a direct comparison to a boundary measurement. We first integrate 0.5 and $1.0\,M_{\rm Jup}$ giants at $q_g=0.04$, 0.05, and 0.06 au. These runs identify 0.05 au as the separation at which changing only the giant mass changes the outcome. We then sample the seven masses listed above. At each mass, a collision-side periapse and an adjacent survival-side periapse separated by 0.005 au bracket the transition. Both endpoints are integrated for three independent orbital-phase realizations, giving 42 boundary integrations and a spatial resolution of 0.005 au.

\subsection{Dynamical integrations}\label{sec:integrations}

We use REBOUND v4.6.0 with the hybrid MERCURIUS integrator and REBOUNDx v4.6.2 for additional forces \citep{ReinLiu2012,Rein2019,Tamayo2020}. MERCURIUS follows the long orbital evolution symplectically and switches treatment during close approaches. The timestep is fixed at $P_{\rm in}/30$, safe mode is enabled, and particles are inserted in Jacobi order. General-relativistic precession is included with the REBOUNDx \texttt{gr} effect. Equilibrium tides and spin evolution are calculated with \texttt{tides\_spin}, which implements the constant-time-lag equations tested by \citet{Lu2023}.

The star, inner planet, and giant have spin periods of 27, 0.5, and 0.5 days. Their Love numbers are 0.07, 0.3, and 0.5, and their moment-of-inertia coefficients are 0.07, 0.33, and 0.25. These choices represent a slowly rotating solar analogue, a terrestrial inner planet, and a gas giant, and retain the physical setup used to establish the Jupiter-mass survival problem \citep{Becker2026}. Their measured radii define the collision surfaces.

A direct integration over a physical gigayear tidal time is computationally prohibitive at the one-day inner-orbit timestep. We therefore rescale the giant's tidal quality factor $Q_g$ in each run so that its nominal circularization time is 1 Myr, and integrate to that time or to the first instability. This acceleration changes the duration of the tidal evolution while retaining the orbital geometry, force model, timestep, and relative sequence of encounters. Because the required $Q_g$ varies with giant mass, imposing $Q_{\rm in}=10^{-3}Q_g$ independently in every run would also change the dissipation inside the inner planet. At each $q_g$, we instead use a common $Q_{\rm in}$ for every mass: the value $10^{-3}Q_g$ obtained for the $1\,M_{\rm Jup}$ case. The fixed-periapse comparison therefore isolates the gravitational effect of $M_g$ and the corresponding outer-planet tide.

We classify an outcome as survival when both planets remain bound at 1 Myr and the inner apocenter lies inside the giant's periapse. A stellar collision, planet--planet collision, or ejection ends the integration and is recorded at its first occurrence. Approximately 2000 output times sample the semimajor axes, eccentricities, periapses, and apsidal separation without calling the integrator at every timestep. At both bracket endpoints for every mass, three deterministic seeds independently draw the inner and outer mean anomalies and the giant's argument of periapse from uniform distributions on $[0,2\pi)$. The masses, orbits, and dissipative parameters are unchanged among these phase realizations.

\subsection{Physical tidal completion time}\label{sec:tcirc}

The accelerated integrations answer whether the inner planet survives the eccentric phase. They do not determine whether the same giant would complete HEM on an astrophysical timescale. We evaluate that second requirement with the unscaled equilibrium-tide circularization time. For tides raised on the giant, the weak-friction equations of \citet{Hut1981}, in the eccentric-planet form used by \citet{Matsumura2008}, give
\begin{equation}
 \tau_{\rm circ}\equiv\frac{e_g}{|\dot e_g|}
 =\frac{1}{27}\frac{Q_p}{n k_{2,p}}
 \frac{M_p}{M_\star}\left(\frac{R_p}{a_g}\right)^{-5}
 \frac{1}{F(e_g,\Omega_p/n)},
 \label{eq:tcirc}
\end{equation}
where
\begin{align}
 F(e,s)={}&\frac{1+\frac{15}{4}e^2+\frac{15}{8}e^4+\frac{5}{64}e^6}
 {(1-e^2)^{13/2}} \nonumber\\
 &-\frac{11}{18}s\,
 \frac{1+\frac{3}{2}e^2+\frac{1}{8}e^4}{(1-e^2)^5}.
 \label{eq:fe}
\end{align}
We evaluate Equations~(\ref{eq:tcirc})--(\ref{eq:fe}) at $a_g=2$ au, with $e_g$ from Equation~(\ref{eq:eout}), $R_p=1\,R_{\rm Jup}$, $k_{2,p}=0.5$, and $\Omega_p/n=1$. These times are separate from the compressed 1 Myr integrations and serve as the physical test of whether migration can finish within 1 or 5 Gyr.

We consider $Q_p=10^5$ and $10^6$. Solar System constraints place giant-planet $Q$ above several times $10^4$ \citep{GoldreichSoter1966}, while population reconstructions of close-in exoplanets permit values extending through $10^6$ and above \citep{Jackson2008}. The adopted range therefore shows how the formation window changes from efficient to weak tidal dissipation. At fixed orbit, stellar properties, radius, and Love number, Equation~(\ref{eq:tcirc}) gives $\tau_{\rm circ}\propto Q_pM_p$.

\section{Results}\label{sec:results}

\subsection{Changing the giant mass changes the fate of the inner planet}\label{sec:masssplit}

\begin{figure*}
\centering
\includegraphics[width=\textwidth]{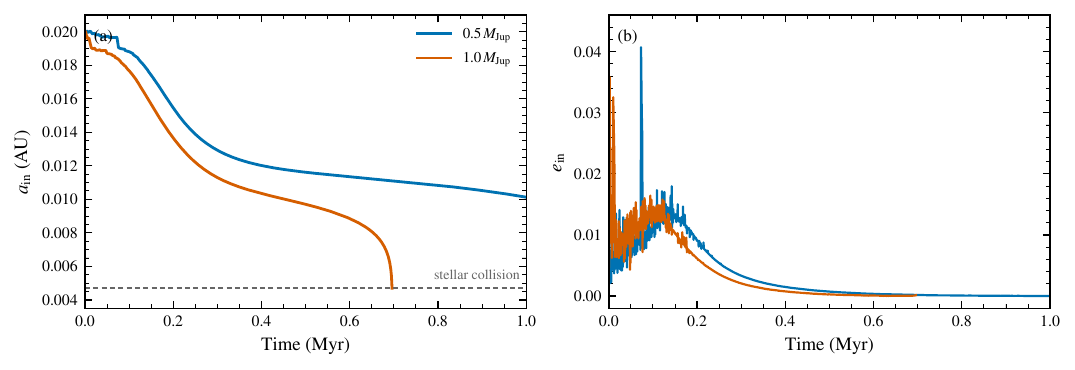}
\caption{Inner-orbit evolution at $q_g=0.05$ au with the same inner-planet dissipation. The inner planet survives the $0.5\,M_{\rm Jup}$ migration and collides with the star during the $1.0\,M_{\rm Jup}$ migration.}
\label{fig:masssplit}
\end{figure*}

Figure~\ref{fig:masssplit} compares 0.5 and $1.0\,M_{\rm Jup}$ giants at the same $q_g=0.05$ au. The $0.5\,M_{\rm Jup}$ case retains the inner planet through 1 Myr; its orbit settles near 0.010 au with negligible eccentricity. The $1.0\,M_{\rm Jup}$ case reaches the stellar surface at 0.696 Myr. Its inner eccentricity remains below 0.05, showing that destruction does not require one extreme eccentricity spike. Repeated modest excitation enables inner-planet tides to remove orbital energy, producing a slow inspiral into the star.

The three phase realizations give the same mass split: all three 0.5$\,M_{\rm Jup}$ runs survive, and all three $1.0\,M_{\rm Jup}$ runs collide at 0.696 Myr. In the surviving case, weaker forcing allows relativistic and tidal precession to decouple the shrinking inner orbit from the giant before it reaches the star. At this periapse, giant mass alone determines whether the system retains evidence of its original inner architecture.

\subsection{The survival boundary moves with giant mass}\label{sec:boundary}

For each giant mass, Table~\ref{tab:boundary} lists the adjacent periapses that bound the change in outcome. The collision--survival boundary moves progressively outward from 0.040--0.045 au at $0.25\,M_{\rm Jup}$ to 0.050--0.055 au by $0.70\,M_{\rm Jup}$. The intermediate masses trace the shift continuously: 0.0425--0.0475 au at $0.40\,M_{\rm Jup}$, 0.045--0.050 au at $0.50\,M_{\rm Jup}$, and 0.0475--0.0525 au at $0.60\,M_{\rm Jup}$. From 0.70 to $1.0\,M_{\rm Jup}$, the boundary is indistinguishable at the 0.005 au sampling resolution.

\begin{deluxetable*}{cccc}
\tablecaption{Outcomes Bounding Inner-Planet Survival\label{tab:boundary}}
\tablehead{
\colhead{$M_g$ ($M_{\rm Jup}$)} &
\colhead{collision-side $q_g$ (au)} &
\colhead{survival-side $q_g$ (au)} &
\colhead{midpoint (au)}
}
\startdata
0.25 & 0.0400 & 0.0450 & 0.0425 \\
0.40 & 0.0425 & 0.0475 & 0.0450 \\
0.50 & 0.0450 & 0.0500 & 0.0475 \\
0.60 & 0.0475 & 0.0525 & 0.0500 \\
0.70 & 0.0500 & 0.0550 & 0.0525 \\
0.85 & 0.0500 & 0.0550 & 0.0525 \\
1.00 & 0.0500 & 0.0550 & 0.0525
\enddata
\tablecomments{Each endpoint was integrated for three independent orbital-phase realizations. All collision-side runs ended in stellar collision, and all survival-side runs remained stable to 1 Myr.}
\end{deluxetable*}

At every mass, all three collision-side realizations collide with the star and all three survival-side realizations remain stable. The full set of 42 boundary integrations therefore contains no mixed outcome at a sampled endpoint. The midpoint Hill spacings decline from $\Delta_H=16.7$ at $0.25\,M_{\rm Jup}$ to 13.1 at $1.0\,M_{\rm Jup}$. Their variation makes $\Delta_H$ an approximate guide whose value changes with mass. The physical boundary retains information about perturbation strength in addition to Hill geometry. The orange envelope in Figure~\ref{fig:window} spans the measured 0.005 au intervals and displays the resolution of the outcome search.

\subsection{Dynamical and tidal boundaries define a viable HEM region}\label{sec:window}

The physical tidal time adds a second limit. At $q_g=0.05$ au and $Q_p=10^6$, $\tau_{\rm circ}$ rises from 5.21 Gyr for $0.25\,M_{\rm Jup}$ to 20.85 Gyr for $1.0\,M_{\rm Jup}$. Equivalently, circularization within 5 Gyr requires $Q_p\lesssim9.6\times10^5$ at $0.25\,M_{\rm Jup}$ and $Q_p\lesssim2.4\times10^5$ at $1.0\,M_{\rm Jup}$, with the intermediate limits scaling inversely with mass. The lower-mass systems that are less destructive dynamically can also finish HEM with weaker tidal dissipation.

\begin{figure*}
\centering
\includegraphics[width=\textwidth]{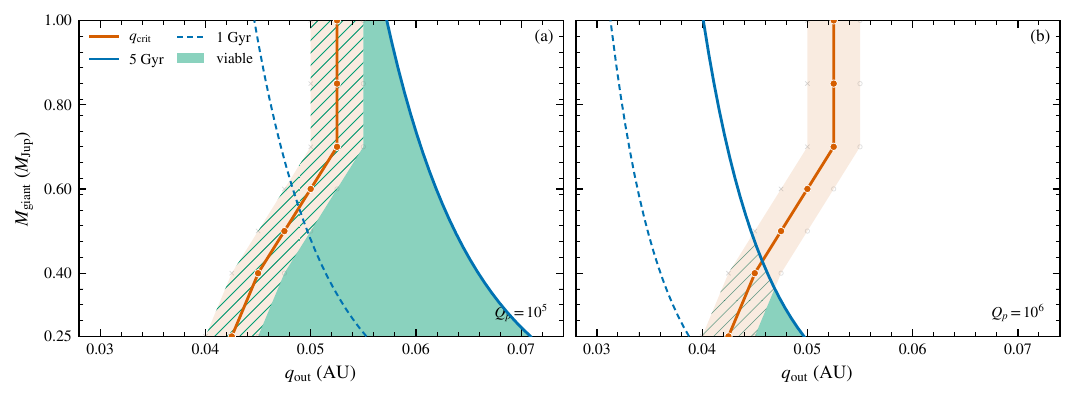}
\caption{Conditions for completing HEM while retaining the inner planet. Orange circles mark the seven sampled boundary midpoints, the line connects them, and the light orange band spans the collision--survival brackets. Faint crosses and open circles mark the collision-side and survival-side endpoints, respectively; each represents three unanimous phase realizations. Blue gives the 5 Gyr (solid) and 1 Gyr (dashed) circularization limits. Green marks survival followed by circularization within 5 Gyr, and hatching marks the part of a sampled bracket compatible with that limit.}
\label{fig:window}
\end{figure*}

Figure~\ref{fig:window} combines the two limits. The orange curve gives the midpoint of the measured collision--survival boundary, and the narrow envelope gives the adjacent outcomes. Survival lies at larger $q_g$. The blue curves mark the periapses at which the giant circularizes in 1 and 5 Gyr; tidal completion lies at smaller $q_g$. Together the boundaries enclose the green viable region: the giant crosses the eccentric phase without destroying the inner planet and subsequently becomes a close-in giant within the adopted age.

For $Q_p=10^5$, the two requirements overlap across the sampled mass range. At $Q_p=10^6$, the overlap contracts toward Saturn masses and periapses near 0.04--0.05 au. A smaller periapse speeds circularization but increases the chance of inner-planet destruction; a larger periapse protects the companion but may leave the giant eccentric for longer than the system age. Giant mass shifts both limits in the same favorable direction at the low-mass end.

\section{Application to Observed Systems}\label{sec:applications}

\begin{figure*}
\centering
\includegraphics[width=\textwidth]{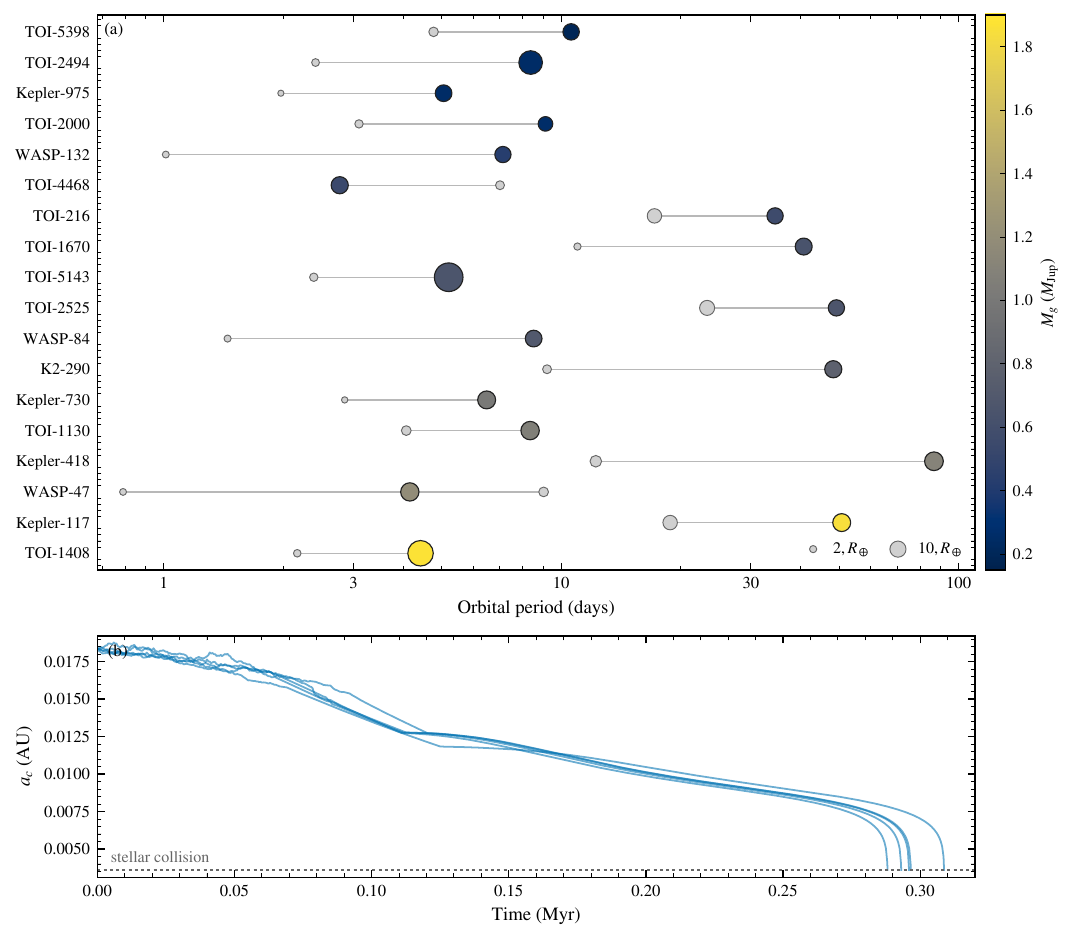}
\caption{Observed companion-hosting architectures and the closest system-specific test. (a) Orbital periods of known nearby-companion systems, ordered by giant mass. Symbol area scales with observed planet radius; colored symbols mark the close-in giant and encode its mass, while gray symbols mark nearby companions. The grazing-transit radii of TOI-2494 c and TOI-5143 c are poorly constrained. (b) WASP-132 c semimajor axis in five full integrations. Every trajectory reaches the stellar collision surface near 0.30 Myr.}
\label{fig:observed}
\end{figure*}

We compiled the current sample from the NASA Exoplanet Archive \citep{Christiansen2025} and recent studies of hot-Jupiter companion systems \citep{Sha2026,Quinn2026,Livesey2026}. Figure~\ref{fig:observed}(a) shows the orbital architectures of 18 companion-hosting hot and warm giants. These systems span a broad range in giant mass despite their broadly Jupiter-like radii, and several lie in the sub-Jovian regime where our calculations predict the largest inward shift of the survival boundary. The complete 16-system inner--giant HEM screen is listed in Appendix Table~\ref{tab:observed}. WASP-47 and TOI-4468 are included in Figure~\ref{fig:observed}(a) for population context but excluded from the pairwise HEM screen: WASP-47 has planets on both sides of the hot Jupiter, whereas TOI-4468 has only an exterior nearby companion.

We then ask whether any of the 16 inner--giant pairs occupies the dynamical--tidal window. For each present-day giant with semimajor axis $a_{g,{\rm now}}$ and eccentricity $e_{g,{\rm now}}$, conservation of orbital angular momentum gives the approximate periapse of a nearly parabolic precursor,
\begin{equation}
 q_{\rm HEM}\simeq\frac{a_{g,{\rm now}}(1-e_{g,{\rm now}}^2)}{2}.
 \label{eq:qhem}
\end{equation}
The survival scale also depends on the observed inner semimajor axis, so Equation~(\ref{eq:hill}) is evaluated separately for every system.

The resulting screen shows that nine reconstructed HEM trajectories cross the present inner orbit. Three additional systems have $\Delta_H<4$, including the low-mass giants TOI-2000 and TOI-2494. Their masses fall in the favorable range identified in Figure~\ref{fig:window}, yet their inner planets are too close to the reconstructed giant periapse. Existing system-specific calculations also remove the inner planets in WASP-84, K2-290, and Kepler-418 \citep{Becker2026}. Giant mass improves the survival prospect, but the inner architecture and migration geometry remain decisive.

WASP-132 is the closest observed test of our reference configuration. The discovery of its inner planet motivated the inference that HEM was unlikely for this system \citep{Hord2022}. Its $0.428\,M_{\rm Jup}$ giant lies near the favorable mass range, and the $6.26\,M_\oplus$ planet c orbits at 0.01833 au \citep{Grieves2025}. The reconstructed $q_{\rm HEM}=0.03369$ au gives $\Delta_H=10.45$, placing the giant well inside the measured survival boundary for this mass.

We integrate the measured $M_\star=0.789\,M_\odot$, $R_\star=0.758\,R_\odot$, $M_c=6.26\,M_\oplus$, $R_c=1.841\,R_\oplus$, $M_b=0.428\,M_{\rm Jup}$, and $R_b=0.901\,R_{\rm Jup}$ for five angular realizations. The giant begins at 2 au with the reconstructed periapse, and $Q_{\rm in}=10^{-3}Q_g$. We omit the distant planet WASP-132 d because the experiment begins after b has reached its eccentric orbit and tests the later interaction between b and c.

All five realizations end in collision between c and the star at 0.288--0.309 Myr (Figure~\ref{fig:observed}(b)). The maximum inner eccentricity ranges from 0.046 to 0.086, while b retains $e_b=0.34$--0.35 when c reaches the star. WASP-132 c is lost through sustained, tide-assisted inspiral before the giant circularizes. Its low giant mass is favorable, but its reconstructed periapse is too small.

No system in the present pairwise sample occupies the viable region. It therefore defines a specific observational prediction: a Saturn-mass close-in giant with a roughly one-day inner planet and a reconstructed HEM periapse near 0.04--0.05 au. Detecting such a system would demonstrate that an inner companion can survive HEM and would provide direct evidence for this formation channel in a compact multiplanet architecture.

\section{Discussion and Conclusions}\label{sec:discussion}

Our calculations identify a mass-dependent survival window for HEM in systems with close inner planets. At the same physical periapse, a Saturn-mass migrant can preserve the inner planet when a Jupiter-mass migrant drives it into the star. Combining the survival boundary with the physical circularization limit isolates a sub-Jovian region where migration can finish without erasing the inner architecture.

At fixed physical periapse, lowering $M_g$ increases the orbital separation in mutual Hill radii because $R_H\propto M_g^{1/3}$ and weakens the perturbation exerted on the inner orbit. The resulting reduction in eccentricity forcing allows relativistic and tidal precession to detach the shrinking inner orbit sooner. The boundary retains the approximate Hill scaling identified for a Jupiter-mass migrator, but the midpoint values $\Delta_H=13$--17 show that no single Hill separation applies across the sampled masses. In the common-radius equilibrium-tide model, lower mass also shortens $\tau_{\rm circ}$ at fixed periapse and $Q_p$. Dynamical survival and tidal completion therefore shift in the same direction, while the contraction of the allowed region at $Q_p=10^6$ shows its sensitivity to giant-planet dissipation.

The slow inspiral in Figure~\ref{fig:masssplit} also connects this survival problem to models for producing ultra-short-period planets. An eccentric exterior giant can maintain a small inner eccentricity and feed tidal orbital decay; moderate excitation may leave an ultra-short-period planet, while stronger excitation drives the inner planet into the star \citep{Zhu2025}. Our collision runs occupy the latter regime.

Planetary multiplicity is therefore a mass-dependent formation diagnostic. HEM generally clears inner systems \citep{Mustill2015,Steffen2012,Sha2026}, but a surviving companion weighs less strongly against HEM when the giant is sub-Jovian. This distinction matters for the observed population because several Jupiter-sized close-in giants have sub-Jovian masses. The absence of a present-day example in the screened sample is consistent with the narrowness of the allowed region. In the current sample, the dominant restrictions are geometric: many systems fail the initial screen, and WASP-132 shows that a favorable mass cannot compensate for a deeply penetrating periapse.

The numerical boundary in Figure~\ref{fig:window} applies to the reference inner orbit at 0.02 au. Changing the inner planet's orbit, mass, radius, or inclination will move it. The physical tidal boundary also depends strongly on the assumed giant radius through $R_p^{-5}$ and on the uncertain $Q_p$. Our two values of $Q_p$ expose this dependence directly. The 0.005 au spacing remains the principal resolution limit. Across all seven sampled giant masses, three independent phase realizations at both bracket endpoints give unanimous classifications, indicating that the measured mass trend is robust to orbital phase at this resolution. Denser sampling in periapse and phase would be required to resolve a narrower chaotic transition and measure a formal survival probability.

Close inner planets therefore provide a mass-dependent test of high-eccentricity migration. Lowering the mass of the migrating giant moves the survival boundary inward and shortens the physical circularization time, opening a sub-Jovian region in which HEM can finish while retaining the inner planet. No presently known system occupies that region. A future detection within it would supply direct evidence that HEM has contributed to the formation of compact, companion-bearing systems.

\begin{acknowledgments}
The author thanks Juliette Becker for helpful discussions, suggestions on the numerical experiments, and comments on the manuscript. This research has made use of the NASA Exoplanet Archive, which is operated by the California Institute of Technology, under contract with the National Aeronautics and Space Administration under the Exoplanet Exploration Program. Generative-AI tools were used to assist with code implementation and language editing. The scientific analysis, interpretation, and conclusions were independently validated by the author, who takes full responsibility for the content of this work.
\end{acknowledgments}

\section*{Data Availability}

Simulation configurations, time series, summary tables, tidal calculations, and analysis scripts are available from a Zenodo repository at \mbox{\href{https://doi.org/10.5281/zenodo.24333223}{10.5281/zenodo.24333223}}.

\software{Astropy \citep{Astropy2022}, Matplotlib \citep{Hunter2007}, NumPy \citep{Harris2020}, pandas \citep{McKinney2010}, REBOUND v4.6.0 \citep{ReinLiu2012,Rein2019}, REBOUNDx v4.6.2 \citep{Tamayo2020,Lu2023}}

\appendix
\renewcommand{\thetable}{A\arabic{table}}
\renewcommand{\theHtable}{A.\arabic{table}}
\setcounter{table}{0}

\section{Observed-system HEM Screen}\label{app:observed}

Table~\ref{tab:observed} gives the pairwise geometric screen summarized in Section~\ref{sec:applications}. The reconstructed periapse follows Equation~(\ref{eq:qhem}), and $\Delta_H$ is evaluated with the measured inner orbit for each system.

\begin{deluxetable*}{lrrrrl}[h]
\tablecaption{Observed Systems in the Pairwise HEM Screen\label{tab:observed}}
\tablehead{
\colhead{System} &
\colhead{$M_g$ ($M_{\rm Jup}$)} &
\colhead{$a_{\rm in}$ (au)} &
\colhead{$q_{\rm HEM}$ (au)} &
\colhead{$\Delta_H$} &
\colhead{HEM screen}
}
\startdata
TOI-1408   & 1.866 & 0.03587 & 0.02889 & $-2.80$ & orbit crossing \\
TOI-216    & 0.560 & 0.12930 & 0.10345 & $-3.50$ & orbit crossing \\
TOI-1130   & 1.057 & 0.04530 & 0.03649 & $-2.75$ & orbit crossing \\
TOI-2525   & 0.657 & 0.15110 & 0.12143 & $-3.34$ & orbit crossing \\
TOI-5398   & 0.185 & 0.05700 & 0.04817 & $-4.25$ & orbit crossing \\
Kepler-730 & 1.000 & 0.03997 & 0.03470 & $-2.09$ & orbit crossing \\
Kepler-117 & 1.840 & 0.14450 & 0.14005 & $-0.38$ & orbit crossing \\
Kepler-975 & 0.252 & 0.02898 & 0.02835 & $-0.50$ & orbit crossing \\
TOI-5143   & 0.632 & 0.03357 & 0.02813 & $-2.85$ & orbit crossing \\
TOI-1670   & 0.630 & 0.10300 & 0.12349 & 3.09 & compact \\
TOI-2000   & 0.257 & 0.04271 & 0.04373 & 0.53 & compact \\
TOI-2494   & 0.258 & 0.03392 & 0.03886 & 2.94 & compact \\
WASP-84    & 0.692 & 0.02359 & 0.03890 & 7.53 & collision\tablenotemark{a} \\
K2-290     & 0.774 & 0.09230 & 0.15250 & 8.20 & collision\tablenotemark{a} \\
Kepler-418 & 1.100 & 0.10632 & 0.18840 & 7.68 & collision\tablenotemark{a} \\
WASP-132   & 0.428 & 0.01833 & 0.03369 & 10.45 & collision (5/5)\tablenotemark{b}
\enddata
\tablenotetext{a}{System-specific integrations reported by \citet{Becker2026}.}
\tablenotetext{b}{System-specific integrations in this work.}
\tablecomments{Negative $\Delta_H$ denotes radial crossing in the reconstructed HEM orbit. ``Compact'' denotes $0<\Delta_H<4$ before a full integration.}
\end{deluxetable*}

\bibliographystyle{aasjournalv7}
\bibliography{references}

\end{document}